\documentclass[aps,notitlepage,prx,onecolumn]{revtex4-2}
\usepackage{listings}
\usepackage{graphicx}
\usepackage{epstopdf}
\usepackage{dcolumn}
\usepackage{bm}
\usepackage{amsmath}
\usepackage{amssymb}

\usepackage{tikz}
\usetikzlibrary{positioning}
\usepackage{subfigure}
\usepackage[export]{adjustbox}

\usepackage{xcolor}

\usepackage{hyperref}
\begin{document}


\title{A Single-Equation Approach to Classifying Neuronal Operational Modes}

\author{Lindsey Knowles}
\author{Cesar Ceballos}%
\author{Rodrigo Pena}
\affiliation{%
 Department of Biological Sciences, Florida Atlantic University, FL 33458, Jupiter, Florida, USA}
\date{\today}

\begin{abstract}
The neural coding is yet to be discovered. The neuronal operational modes that arise with fixed inputs but with varying degrees of stimulation help to elucidate their coding properties. In neurons receiving {\it in vivo} stimulation, we show that two operation modes can be described with simplified models: the coincidence detection mode and the integration mode. Our derivations include a simplified polynomial model with non-linear coefficients ($\beta_i$) that capture the subthreshold dynamics of these modes of operation. The resulting model can explain these transitions with the sign and size of the smallest nonlinear coefficient of the polynomial alone. Defining neuronal operational modes provides insight into the processing and transmission of information through electrical currents. Requisite operational modes for proper neuronal functioning may explain disorders involving dysfunction of electrophysiological behavior, such as channelopathies.
\end{abstract}
\maketitle
\section{Introduction}
Neurons are widely understood to be the fundamental computational units of the brain and operate in qualitatively different modes depending on their dynamics \cite{KocSeg00}. A well-known distinction is made between coincidence detectors and integrators \cite{Kanev2016,Bre15}. Coincidence detectors are sensitive to the precise firing of inputs, incoming signals, and fire only when pulses arrive in close synchrony. Integrator neurons, by contrast, act as accumulators, summing inputs over time until a threshold is reached and a spike is produced. Neurons exist on a continuum between these modes, and thus both of these two modes play crucial roles in shaping how neurons communicate information \cite{Rudolph2003}.

Traditionally, identifying whether a neuron behaves as an integrator or a coincidence detector has relied on direct experimentation. Previous attempts to characterize operational modes computationally required modeling precisely timed synaptic inputs across wide parameter ranges, making these models high-dimensional and computationally expensive \cite{Kanev2016}. Here, we explore how reducing the Hodgkin-Huxley (HH) mathematical model for the initiation and propagation of action potentials in neurons to a simplified polynomial model can provide a complementary perspective. By treating neuronal firing as a bifurcation problem in a one-dimensional dynamical system, we can analyze how changes in parameters shift the system between detector and integrative modes.

\begin{figure}[h]%
    \centering
    \subfigure[]{\includegraphics[width=0.27\linewidth]{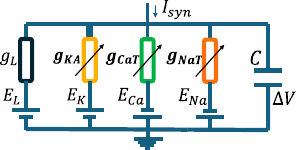} }%
    \smash{\Huge{$\Rightarrow$}}
    \subfigure[]{{\includegraphics[width=0.3\linewidth]{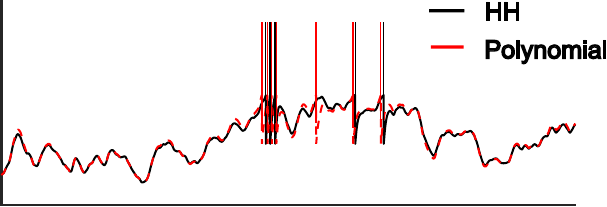}}}%
    \smash{\Huge{$\Rightarrow$}}
    \subfigure[]{{\includegraphics[width=0.25\linewidth]{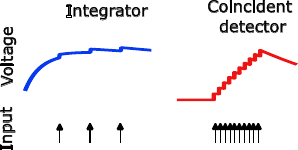} }}%
    \caption{The problem addressed in this paper. (a) We simplify Hodgkin-Huxley neuron models to (b) polynomial models that accommodate different (c) integration modes under {\it in vivo} stimulation.  Simulations show that the simplified model almost perfectly matches the Hodgkin-Huxley version.}%
    \label{abs}%
\end{figure}

As outlined in Figure \ref{abs}, this study addresses the identification of neuronal operational modes through simulation. Specifically, we reduce the HH model to a single polynomial equation, allowing us to identify the integration mode of a neuron computationally.

\section{Methods}
\subsection{\label{2.1}Neuron Model}
In our neuron model, we consider a single compartment with membrane voltage described by the equation
\begin{align}
    C \frac{dV}{dt} = -I_L + I_i,
\end{align}
where $I_i$ contains the $i\text{th}$ ionic currents following
\begin{align}
    I_i = g m h (V-E_\mathrm{rest}),
\end{align}
where $g$ is the conductance, $E_\text{rest}$ is the reversal potential, $m$ is activation, and $h$ is inactivation \cite{HodHux52}. 

For $x=(m,h)$, we have 
\begin{align}
\frac{dx}{dt}=\frac{x^\infty-x}{\tau_x}. 
\end{align}

The equation of the steady-state activation/inactivation follows a sigmoidal expression as
\begin{equation}
    x^\infty=\frac{1}{1+\exp({\frac{V-V_{1/2}}{k}})},
    \label{Eq:sig}
\end{equation}

\noindent where $V_{1/2}$ represents the voltage in which $x^{\infty}= 0.5$ and $k$ is the slope of $x^{\infty}$ function.





\subsection{\label{2.4}System, Equilibria, and Stability Analysis}

In the polynomial model we have thus far, an important consideration is balancing the value of $a_0$ to ensure that $a_0+\sum_{k=1}^n a_kV^k=0$ at rest. However, by centering on the origin, we can analyze system dynamics where the $a_0$ value is set to $0$.

We begin with equation $\frac{dV}{dt}=P(V)+I(t)$ where 
\begin{equation}
    P(V)=\sum_{k=0}^na_kV^k.
\end{equation}
If we shift the variables by their deviation from rest so $X:=V-E_\mathrm{rest}$ we can use Taylor expansion (\ref{Texp}) to rewrite the drift in powers of $X$ as
\begin{equation}
    Q(X)=\sum_{i=0}^nc_iX^i,
\end{equation}
where $c_i=\frac{P^iE_\mathrm{rest}}{i!}$ and $P(E_\mathrm{rest})=0\implies c_0=0$. Here, $c_1$ is the slope at rest (leak) and $c_2,\dots,c_n$ are local nonlinearities.

We further simplify the equation to remove physical dimensions from the model so that the coefficients are dimensionless and comparable \cite{holmes}.
Let the voltage scale be represented by $\Delta V:=V_\text{th}-E_\text{rest}$ and let the time scale $\tau:=1/|c_1|$ be the linear relaxation time near rest. We can now define dimensionless variables $y:=\frac{X}{\Delta V}\in [0,1]$ and $s:=\frac{t}{\tau}$ and apply the chain rule (\ref{Nondimen}) to represent the change in voltage as
\begin{equation}
    \frac{dy}{ds}=R(y)+\eta(s),  
\end{equation}
where $R(y)=\sum_{i=1}^n\beta_iy^i$, $\beta_i:=c_i\tau(\Delta V)^{i-1}$, and $\eta:=\frac{I\tau}{\Delta V}$. $\beta_1$ will now be set to $\pm1$, which sets the time constant to $1$ and $\beta_i$ are now dimensionless and comparable. 

From the derived model, we study a one-dimensional polynomial dynamical system of the form
\begin{equation}
    \dot{y}=f(y)=R(y)+\eta(s),
\end{equation}
where $y\in \mathbb{R}$ is the state variable and coefficients $\beta_i \in \mathbb{R}$ are system parameters representing the dynamics of the equation \ref{Nondimen}. Here we study the subthreshold dynamics of the neurons within the range [0,1], where $y = 0$ represents the voltage at rest and $y = 1$ represents the firing threshold. 

$\beta_i$ represents the leak current, and is fixed to $\pm1$ by definition \ref{Nondimen}. Stable neurons have a negative leak coefficient while positive leak indicates the presence of a drive or bias injected current. Hence, the sign of $\beta_1$ indicates stability.

Equilibria are defined by the real roots of $f(y)=0$ \cite{Str24}. The stability of each equilibrium $y$ is determined by the sign of the derivative $f'(y)$. If $f'(y)<0$, the fixed point is stable (attracting), whereas if $f'(y)>0$, it is unstable (repelling).

Suppose $m=\min\{j\geq2:\beta_i\neq0\}$. Assuming a stable, negative leak, if $\beta_m < 0$, then $f'(y)<0$, indicating that the system acts as a coincidence detector being pulled to zero, and an input current $\eta(s)$ must be large enough to counter the system and initiate a spike.  Alternatively, under the same stable leak condition, if $\beta_m > 0$, $f'(y)>0$, then there exists a repellor near rest, allowing the system to accumulate charge until it reaches threshold as an integrator.

\begin{figure}[ht]%
    \centering
    \subfigure[]{{\includegraphics[width=0.4\linewidth]{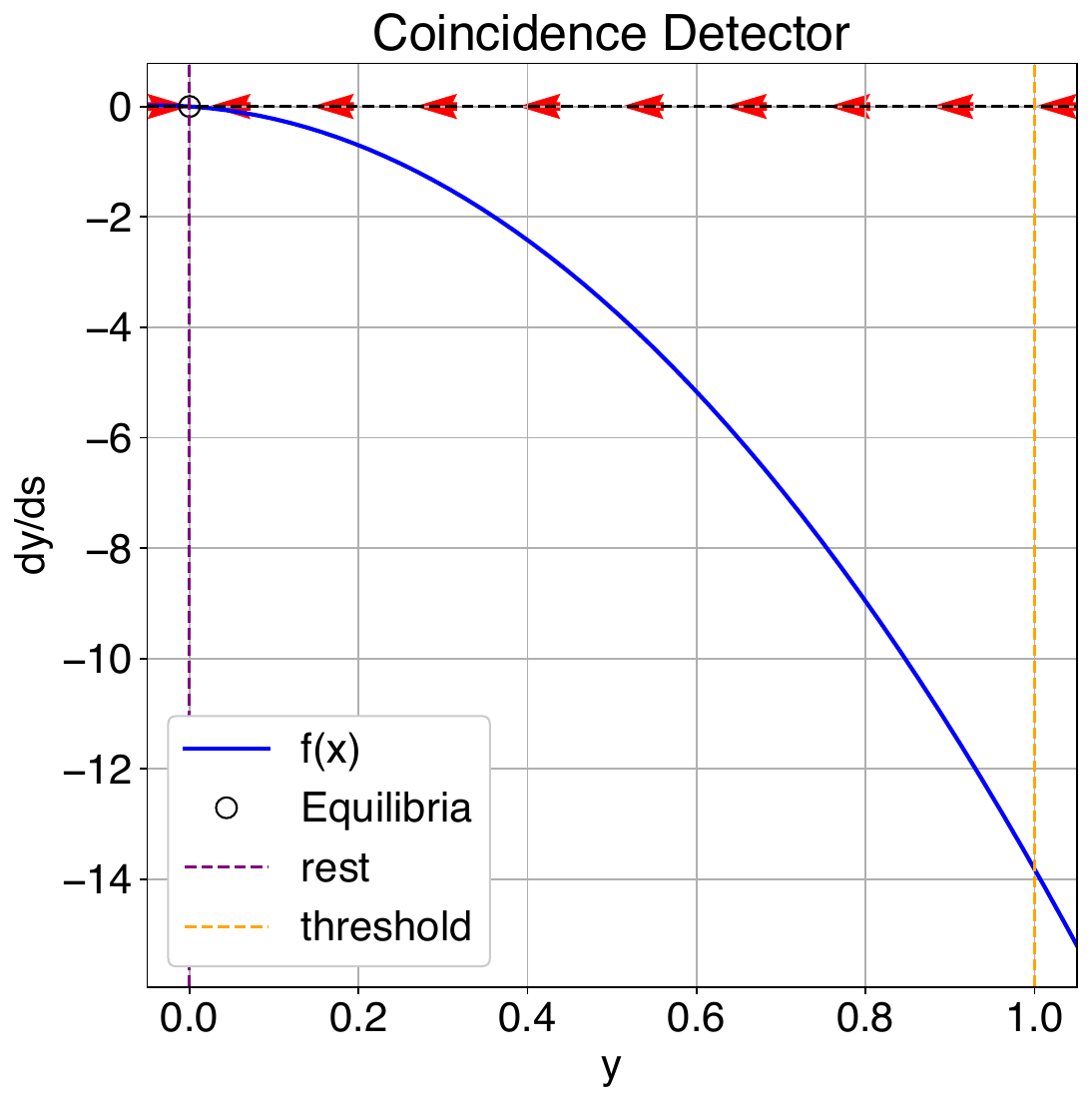} }}%
    \qquad
    \subfigure[]{{\includegraphics[width=0.4\linewidth]{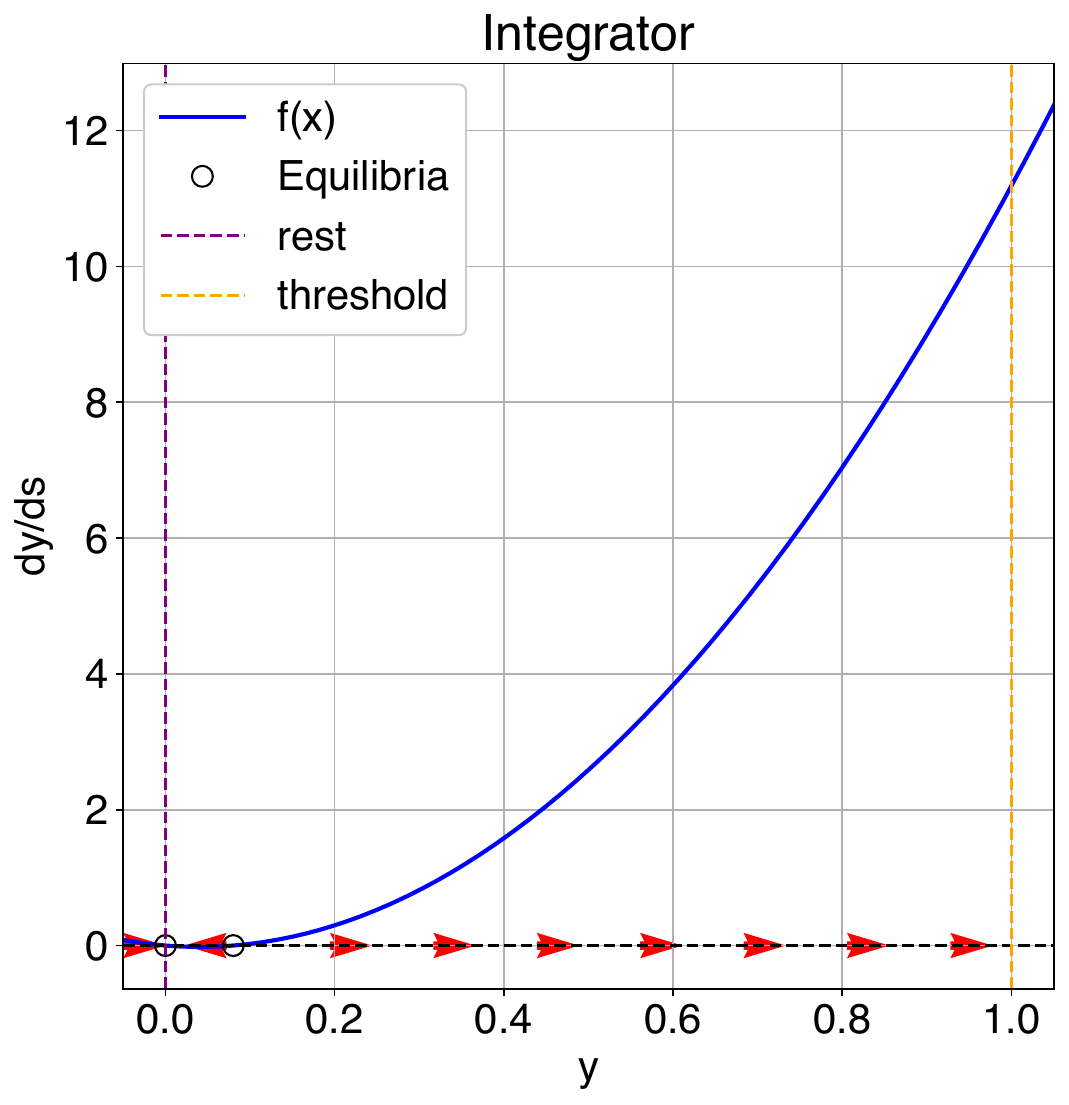} }}%
    \caption{Firing patterns of cubic polynomial neurons with negative leak terms. (a) Detector operational mode system with stable, negative leak coefficient ($-\beta_i$) and negative $\beta_m$ ($\beta_2$).  (b) Integrator operational mode system with stable, negative leak coefficient ($-\beta_i$) and positive $\beta_m$ ($\beta_2$).}%
    \label{cdi}%
\end{figure}

Phase portraits depicted in Figure \ref{cdi} confirmed that with a stable negative leak coefficient, trajectories with a negative $\beta_m$ pull the system back to rest after firing. Since the trajectory faces slow velocity until it reaches the threshold, any presynaptic spike will cause a long-lasting depolarization to the membrane of a neuron, indicating detector operational mode. Meanwhile, positive $\beta_m$ or positive provide equilibria between rest and threshold, pushing the system to threshold through integration.

Flipping the sign on the linear term perturbs the polynomial and shifts all equilibrium points. Unlike the quadratic case, higher-order systems do not generally exhibit simple symmetry: the roots move in a nonlinear manner determined by the remaining coefficients. Stability may change only if the parameter change causes equilibria to collide or split (i.e., when the system crosses a bifurcation point). Otherwise, the qualitative stability structure is preserved.

Increasing the magnitude of the lowest-order nonlinear coefficient strengthens curvature in the vector field, amplifying deviations from linearity. This broadens the separation of equilibria and can alter their stability through changes in the local slope $f'(y)$.  Conversely, reducing this coefficient flattens the system toward linear behavior, diminishing nonlinear effects and potential bifurcations. Therefore, the magnitude $|\beta_m|$ indicates how strongly the neuron will behave as either a coincidence detector or integrator.

\section{Results}

\subsection{Simplified polynomial neuron model reproduces modes of integration} 

Under in-vivo like synaptic input, we can assume that $\tau_m \rightarrow 0$ and $\tau_h \rightarrow0$ \cite{CebPen25}. Thus, $m=m^\infty$ and $h=h^\infty$, and the new current equation is 
\begin{align}
I=gm^\infty h^\infty(V-E_\text{rest}).
\end{align}

We find from phase plots comparing activation against inactivation of the variable's trajectories under in-vivo like simulation that we can approximate $m^\infty$ as the quadratic function $m^\infty=a_2h^{\infty^{2}}+a_1h^\infty+a_0$. Also, we can approximate the sigmoid of Eq.~\ref{Eq:sig} within the voltage interval observed under the in-vivo fluctuation as a quadratic function: $h^\infty=b_2V^2+b_1V+b_0$ ($R^2$ close to 1). This suggests that a general fit corresponds to a polynomial. So then we can approximate $m^\infty=\sum_{k=1}^{n}a_k(V)^k$ where $k\in \mathbb{Z}^+$. This allows us to rewrite the current equation as \begin{eqnarray}
    I &=& \sum_{k=0}^{n}a_k(V)^{k+1}(V-E_\mathrm{rest})
\end{eqnarray}

\begin{figure}[ht]%
    \centering
    \subfigure[]{{\includegraphics[width=0.4\linewidth]{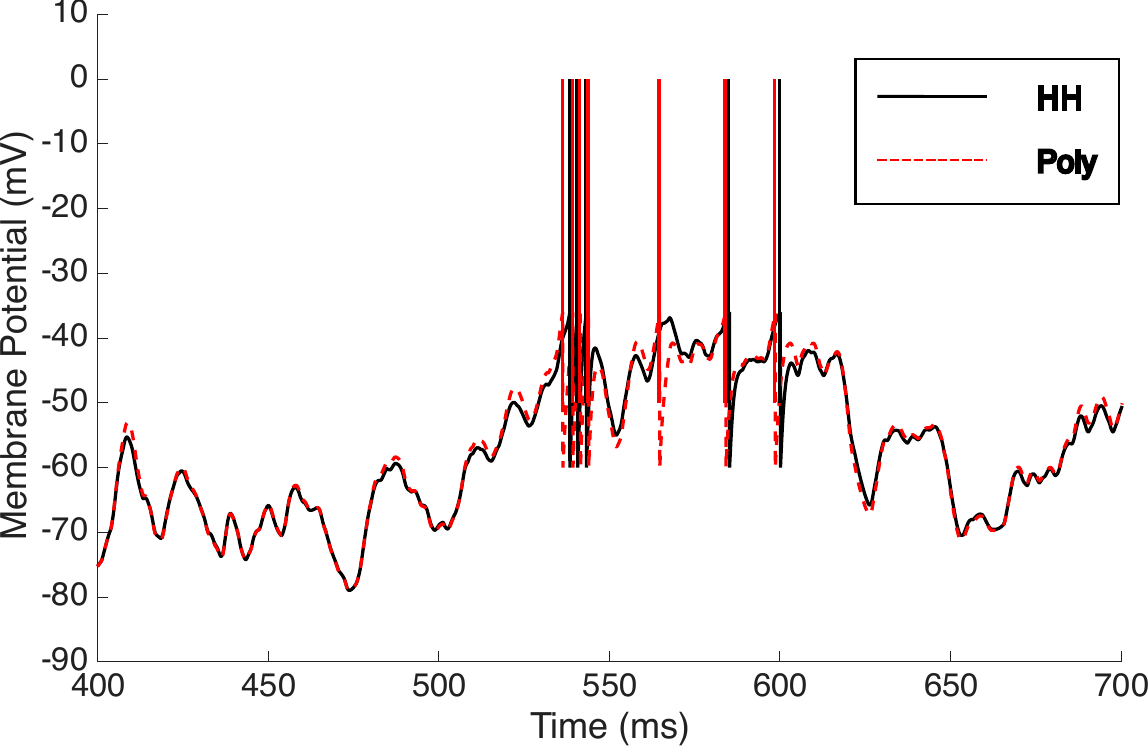} }}%
    \qquad
    \subfigure[]{{\includegraphics[width=0.4\linewidth]{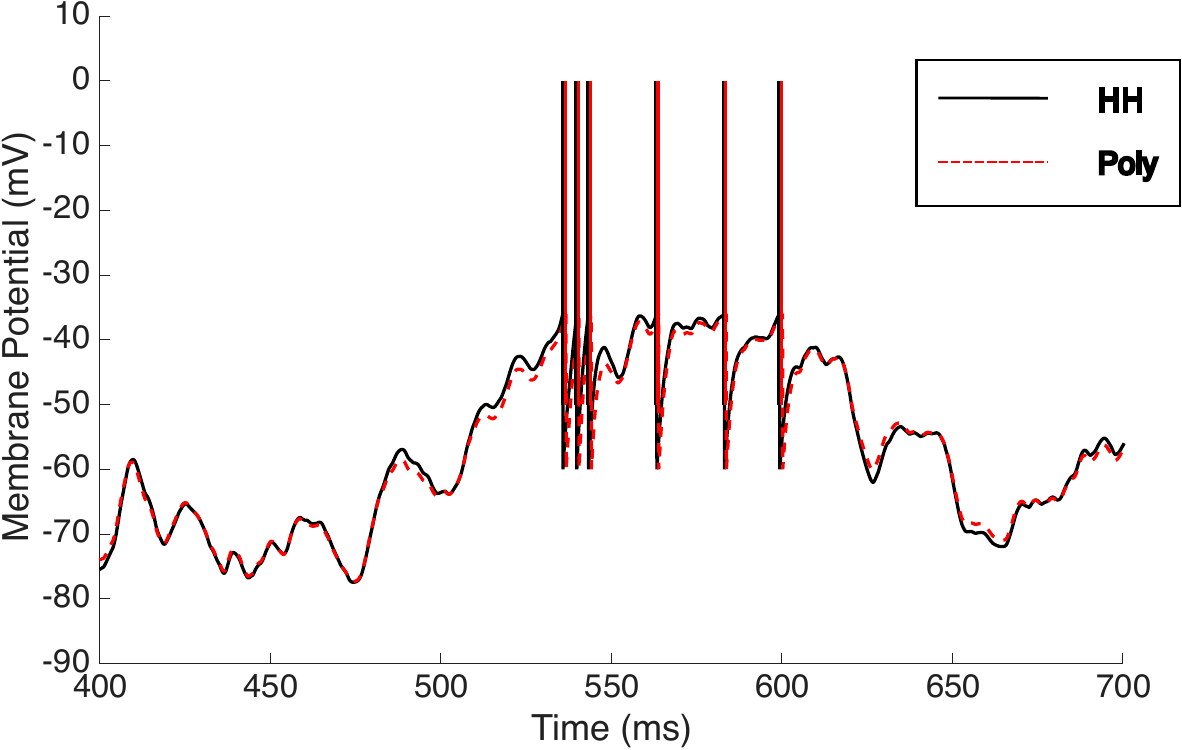} }}%
    \caption{Accuracy of the polynomial model compared to the Hodgkin-Huxley model for neuronal dynamics of (a) potassium and (b) sodium channels.}%
    \label{fit}%
\end{figure}

Substituting this into the current equation, we solve for $I$ and get
\begin{equation}
    I=\sum_{k=0}^{2n+3}a_kV^k.
\end{equation}
As shown in Figure \ref{fit}, we can fit the coefficients of the polynomial model to match the HH model's dynamics \ref{HHpoly}.

\begin{figure}[ht]%
    \centering
    \subfigure[]{{\includegraphics[width=0.4\linewidth]{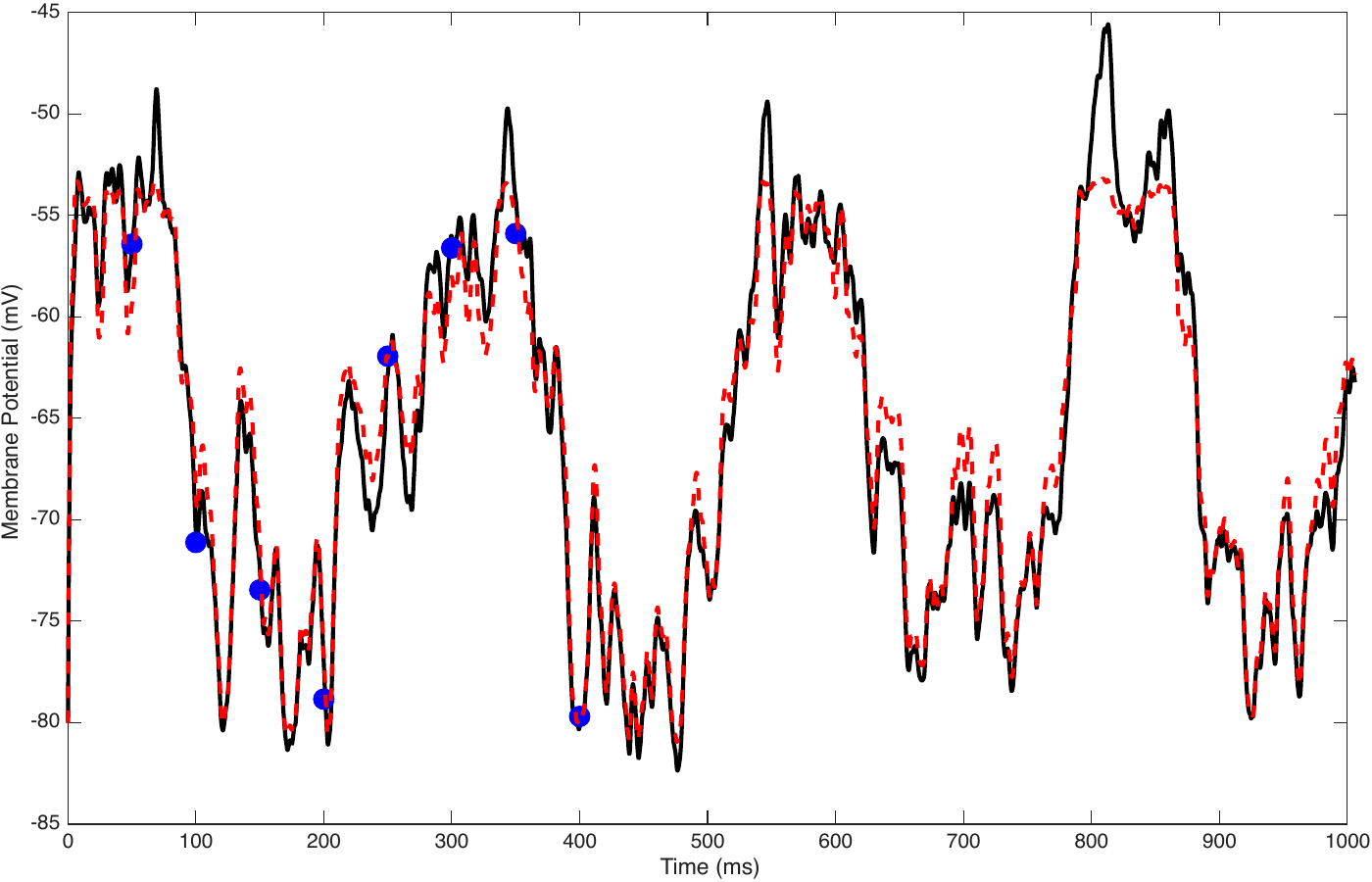} }}%
    \qquad
    \subfigure[]{{\includegraphics[width=0.4\linewidth]{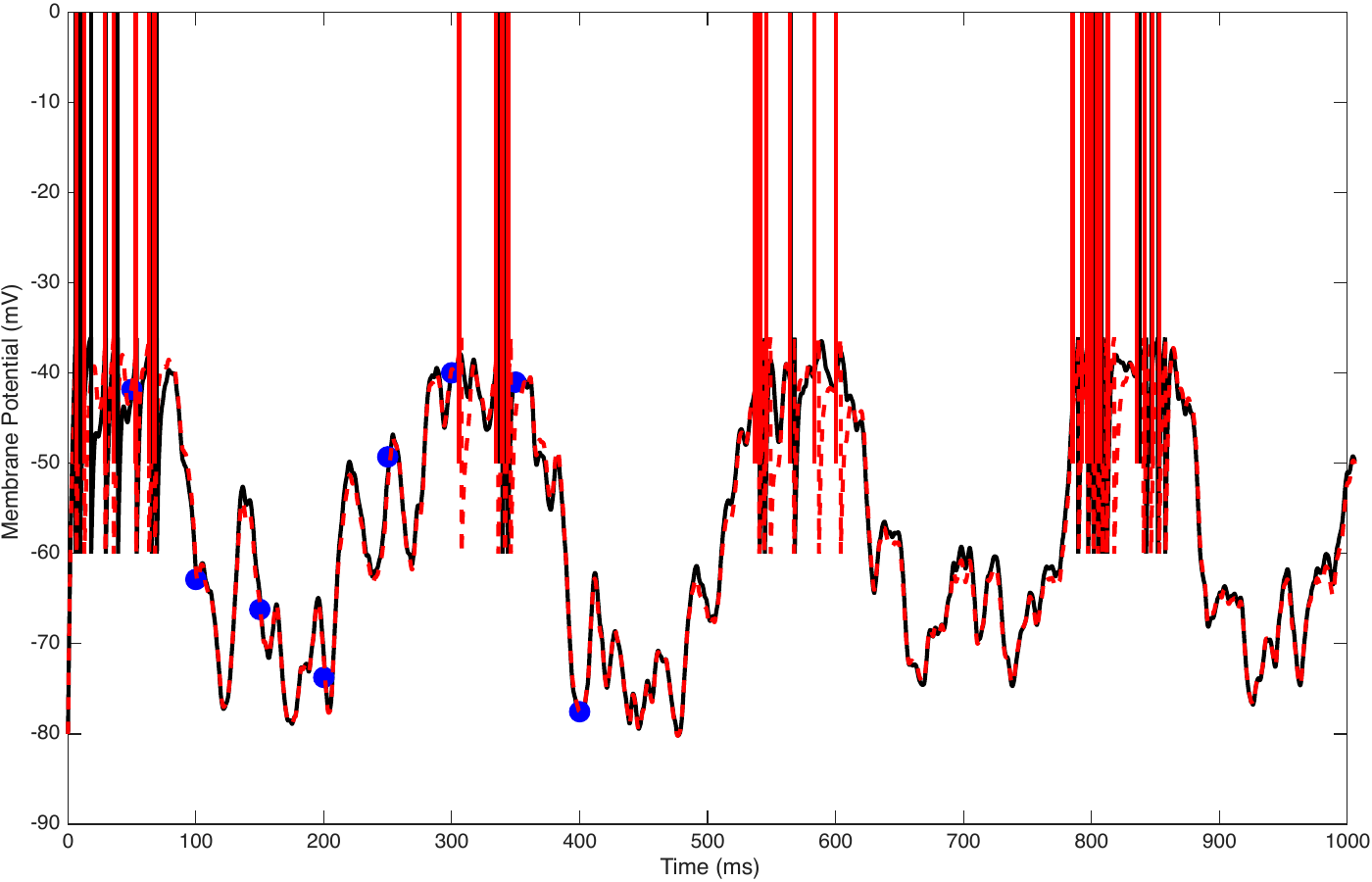} }}%
    \caption{For these polynomials of order 7, the model is able to accurately predict the trajectories of the membrane potentials from the HH models for (a) potassium and (b) sodium channels..}%
    \label{rev}%
\end{figure}

With $k+1$ given points, the model not only fits the HH model but also accurately predicts the trajectory of the membrane potential over time, as seen in Figure \ref{rev}. 



\subsection{\label{system}Dynamical Systems Analysis Explains the Transition Between Operation Modes}



\begin{figure}[!ht]%
    \centering
    \subfigure[]{{\includegraphics[width=0.4\linewidth]{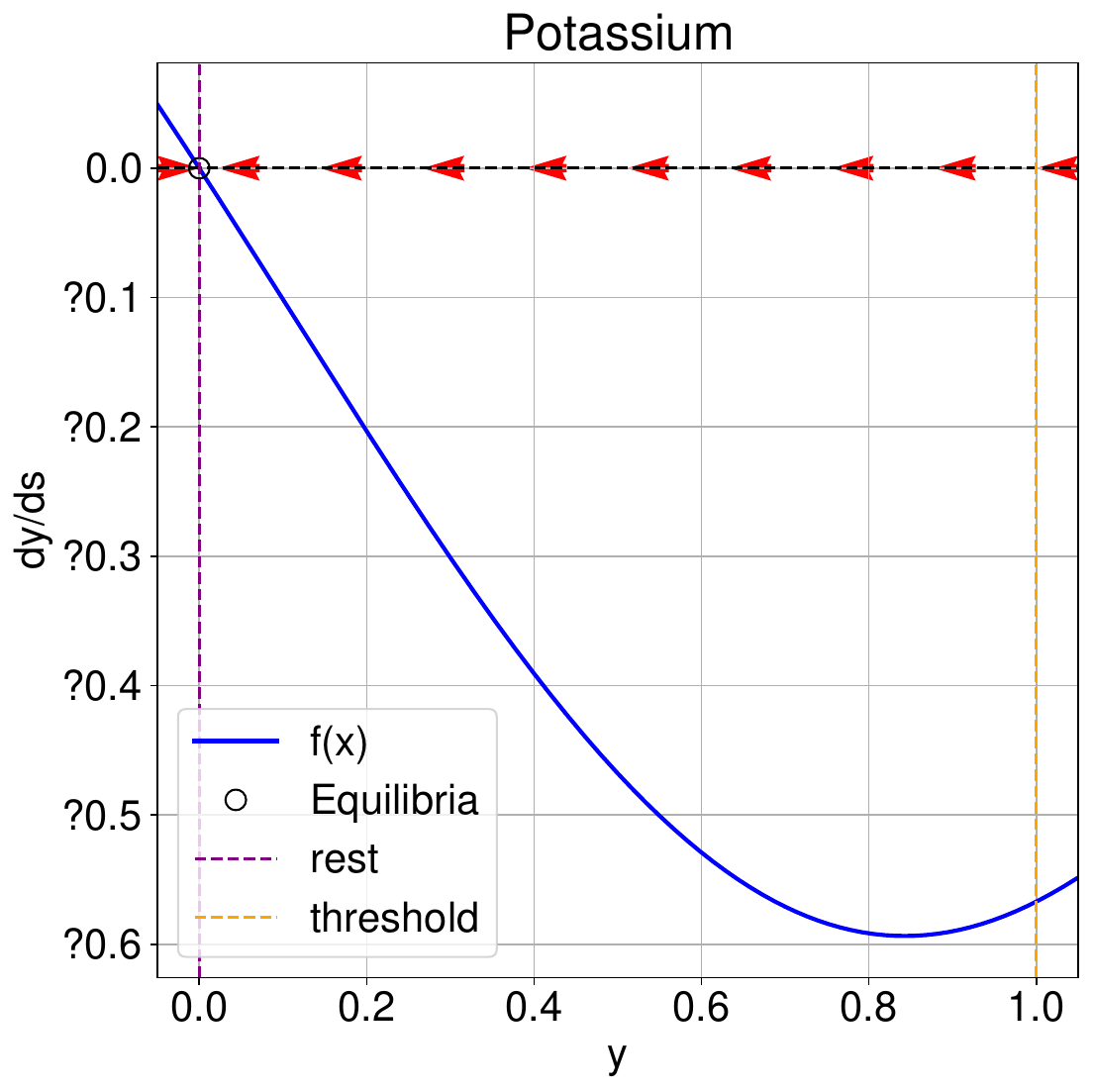} }}%
    \qquad
    \subfigure[]{{\includegraphics[width=0.4\linewidth]{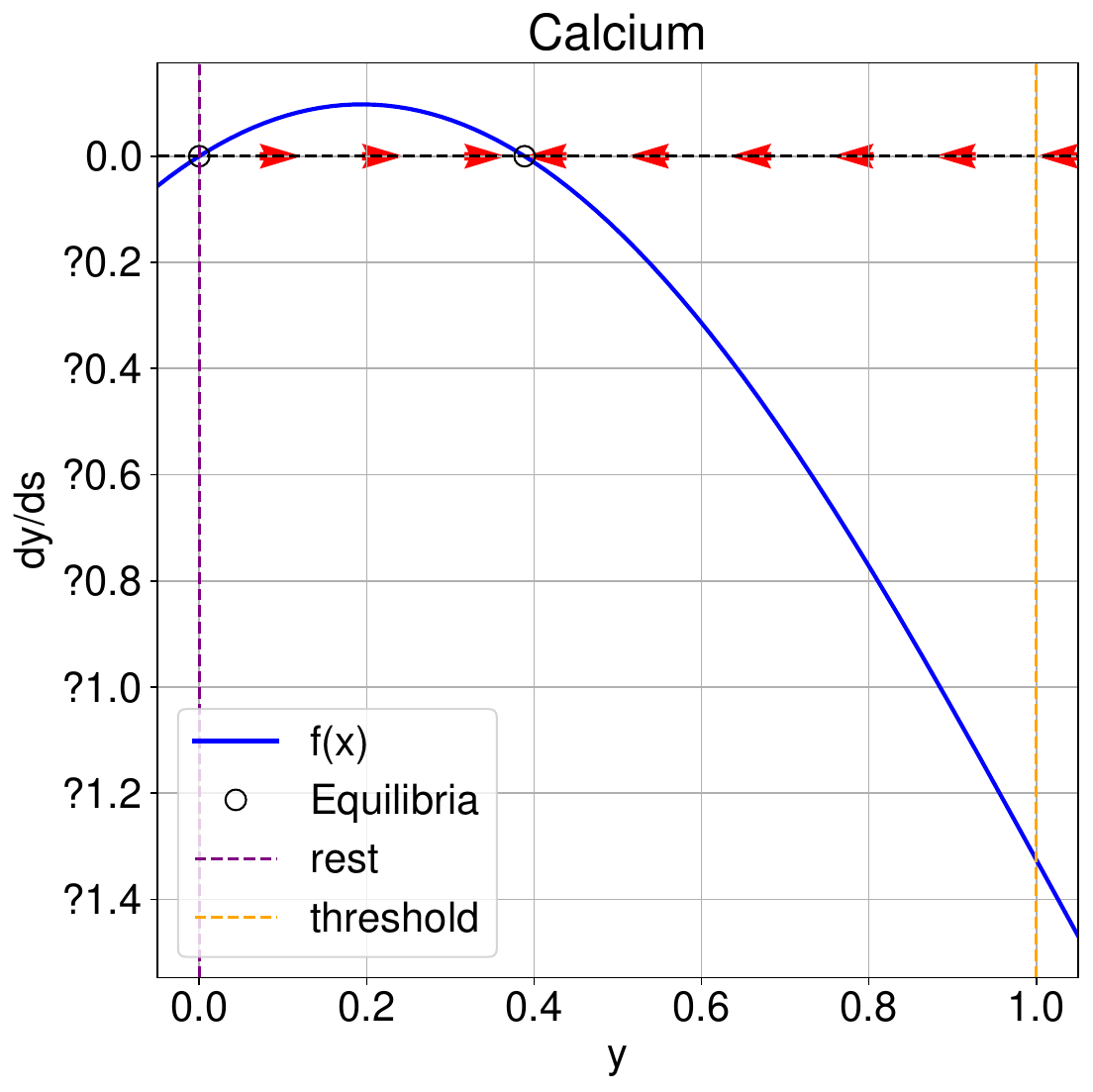} }}%
    \caption{Phase portraits for polynomials representing potassium and calcium. (a) The potassium channel's stable, negative leak term allows the dynamics of the system to reflect the single-equation approach. (b) A ``postive leak" (bias current injection) shifts the equilibria of calcium's equation, allowing for the channel's integrator operational mode despite the equation's $-\beta_m$ value. }%
    \label{pot_cal}
\end{figure}

We center and scale the polynomials representing potassium and calcium channels, giving us their respective polynomial equations \ref{channel.simple}. As seen in Figure \ref{pot_cal}, both equations have negative $\beta_m$ terms; however, the calcium channel has a positive leak term while the potassium channel presents a negative leak term. The positive $\beta_1>0$ indicates not a leak but rather a bias or drive input current and indicates instability. This shifts calcium's polynomial north, providing it with an attractor between rest and threshold and indicating integrator dynamics.


\section{\label{sec:level1}Discussion}
In this paper, we reduce the Hodgkin-Huxley model to a simplified, centered, nondimensionalized polynomial equation that faithfully reproduces different operational modes under {\it in vivo} situations. The resulting one-dimensional dynamical system based on polynomial behavior switches modes mostly from the quadratic equation coefficient.

Coincidence detector neurons fire when a series of pulses arrive simultaneously, and conversely, integrators accumulate inputs to reach an objective threshold. \cite{Izh07,Bre15,RatHon13,HonRat12,StoMai}. The two-dimensional dynamical system that emerges from the polynomial model provides an approach for identifying neuronal operational modes based on subthreshold equilibria, with subthreshold repellors allowing integrator neurons to accumulate charge until threshold and subthreshold attractors pulling the detector system to rest and hence requiring a strong, synchronous current injection to push the system to threshold and initiate an action potential. 


Our model shows that positive values for the lowest order nonlinear coefficient $\beta_m$ tended to behave as integrators, while negative values tended to behave as coincidence detectors. Furthermore, as the magnitude of the lowest–order nonlinear coefficient $|\beta_m|$ increases, the equilibria separate more distinctly in state space, which sharpens the local gradients around them, strengthening the tendency towards the operational mode associated with $\beta_m$'s sign.


From a physiological standpoint, the quadratic coefficient bundles together several biophysical factors, including the steepness of activation curves, resting potentials and, implicitly, neuromodulatory state. Small mutations or pharmacological manipulations that shift those factors can thus move a neuron across the boundary that were mapped. The framework offered here makes two concrete, experimentally testable predictions: (i) interventions that flatten the sub-threshold I-V curve near rest should convert coincidence detectors into integrators, and vice-versa; (ii) neurons whose I-V curves exhibit opposite curvature but share similar linear leak should respond differently to synchronous versus diffuse synaptic drive, a difference that can be quantified with the relative-frequency index used in our simulations.

Admittedly, the polynomial reduction is deliberatively minimal and therefore has limitations. It ignores spike-frequency adaptation \cite{BenHer03}, dendritic filtering \cite{CebPen25,DasWhi07} and the interaction between compartments \cite{GolRem19}, all of which can influence operational mode in real neurons. Nevertheless, its analytical tractability enables rapid parameter sweeps and mechanistic insight that would be difficult to obtain from high-dimensional conductance‐based models. Extending the present approach to include an adaptation variable or a second spatial compartment, while retaining the single-equation spirit, is a promising avenue for future work.

In summary, the results in this paper show how simple polynomial nonlinearities capture subthreshold dynamics to understand the typical operational behavior of neurons in a system. With these approximations, we may still learn how typical operational modes against atypical cases like the ones seen in channelopathies occur in terms of the nature of ion channel dysfunctions.

\appendix

\section{Polynomial Model Simplification}
\subsection{\label{Texp}Taylor Expansion}
Let the voltage-drift polynomial be
\begin{equation*}
    P(V) = \sum_{k=0}^n a_kV^k.
\end{equation*}
To obtain a representation centered at the resting membrane potential $E_\mathrm{rest}$, define the shifted voltage coordinate
\begin{equation*}
    X:=V-E_\mathrm{rest}.
\end{equation*}
Then $P(V) = P(X + E_\mathrm{rest})$ admits the Taylor expansion
\begin{equation*}
    P(V) = \sum_{i=0}^n c_iX^i, \quad \text{with} \quad c_i = \frac{1}{i!}P^{(i)}(E_\mathrm{rest}).
\end{equation*}
Explicitly,
\begin{eqnarray*}
    c_0&=&P(E_{\mathrm{rest}})\\   c_1&=&P'(E_{\mathrm{rest}})=\sum_{k=1}^nka_kE_{\mathrm{rest}}^{k-1} \nonumber\\
    c_2&=&\frac{1}{2}P''(E_{\mathrm{rest}})=\sum_{k=1}^n\frac{k(k-1)}{2}a_kE_{\mathrm{rest}}^{k-2}\\
    \vdots \nonumber\\
    c_n&=&\frac{1}{n!}P^{(n)}(E_{\mathrm{rest}})=a_n
\end{eqnarray*}
If $E_\mathrm{rest}$ is an equilibrium point of the dynamics, then $P(E_\mathrm{rest})=0$, so $c_0=0$ and the rest-centered drift polynomial takes the form
\begin{equation*}
    Q(X)=c_iX^i.
\end{equation*}

\subsection{\label{Nondimen}Chain Rule for Nondimensionalization}

To nondimensionalize voltage, define
\begin{equation*}
    y:=\frac{X}{\Delta V}, \quad \Delta V:=V_\mathrm{th}-E_\mathrm{rest}>0.
\end{equation*}
Then, by the chain rule,
\begin{equation*}
    \frac{dy}{dt}=\frac{1}{\Delta V}\frac{dX}{dt}.
\end{equation*}
We also introduce a dimensionless time variable
\begin{equation*}
    s:=\frac{t}{\tau}, \implies \frac{dt}{ds} = \tau,
\end{equation*}
so that
\begin{equation*}
    \frac{dy}{ds}=\frac{dy}{dt} \frac{dt}{ds}=\frac{\tau}{\Delta V}\frac{dX}{dt}
\end{equation*}
Using the rest-centered polynomial
\begin{equation*}
    \frac{dX}{dt}=\sum_{i=1}^n c_iX^i + I(t),
\end{equation*}
we obtain
\begin{equation*}
    \frac{dy}{ds} =\frac{\tau}{\Delta V} \left( \sum_{i=1}^n c_iX^i + I(t)\right).
\end{equation*}
Substituting $X= \Delta V y$ gives
\begin{equation*}
    \frac{dy}{ds} = \sum_{i=1}^{n}c_i\tau (\Delta V)^{i-1} y^i+\frac{\tau}{\Delta V}I(t)
\end{equation*}
We therefore define dimensionless coefficients and input
\begin{equation*}
    \boxed{\beta_i=c_i\tau(\Delta V)^{i-1}, \quad \eta(s)=\frac{\tau}{\Delta V}I(t),}
\end{equation*}
yielding the reduced system
\begin{equation*}
    \frac{dy}{ds}=\sum_{i=1}^n \beta_i y^i + \eta(s).
\end{equation*}

\subsection{\label{HHpoly} Potassium and Calcium Current Modeling}
A single-compartment neuron model was implemented in MATLAB to evaluate conductance-based representations of voltage-gated ionic currents. The model compares two formulations—one using standard Hodgkin–Huxley gating and another using a fitted polynomial surrogate—for both A-type potassium ($\text{K}_\mathrm{A}$) and calcium ($\text{Ca}^{2+}$ currents.

The membrane potential was updated by 
\begin{equation*}
    C_m \frac{dV}{dt} = g_L(E_L-V) + I_\mathrm{ion} + I_\mathrm{inj},
\end{equation*}
with $C_m =1\, \mathrm{\mu F/cm^2}$, $g_L =0.3\, \mathrm{mS/cm^2}$, $E_L =-80\, \mathrm{mV}$.
For potassium conditions, $g_A =1\, \mathrm{mS/cm^2}$ and $E_K = -80\, \mathrm{mV}$; for calcium, $g_A =0.1\, \mathrm{mS/cm^2}$ and $E_{Ca} = +80\, \mathrm{mV}$. Full activation/inactivation dynamics ($n,\,l$) were simulated with
\begin{equation*}
    \frac{dn}{dt} = \frac{n_\infty(V) - n}{\tau_n}, \quad \frac{dl}{dt} = \frac{l_\infty (V) -l}{\tau_l},
\end{equation*}
where $n_\infty (V)$ and $l_\infty (V)$ are sigmoidal functions of voltage. A sixth-order polynomial fit of $n_\infty (V) \_\infty (V)$ replaced the gating variables, giving a direct voltage-dependent current expression.

Both channel types were driven by a $4 \,\mathrm{Hz}$ sinusoidal current plus background synaptic fluctuations derived from recorded data. The surrogate model replicated the spiking and subthreshold behavior of the full formulation while significantly reducing computational cost.

\subsection{\label{channel.simple}Potassium and Calcium Channel Simplifications}

To support fast reduced-order neuron modeling, the fitted voltage-polynomial representations of the ionic currents (derived separately for the A-type potassium and calcium channel cases) were transformed into a rest-centered and dimensionless form. This allows direct comparison across channel types and operating regimes while avoiding numerical sensitivity to membrane-potential offsets and physical units.

The constant term of the fitted drift polynomial $P(V)$ is adjusted so that $P(V_\mathrm{rest})=0$, ensuring the true resting potential is a fixed point of the reduced dynamics.
Then, the polynomial is re-expressed around $X = V-E_\mathrm{rest}$, so that $X=0$ corresponds to rest rather than absolute voltage. This yields a polynomial $Q(u)$ that is numerically well-conditioned near rest.
Voltage and time are rescaled using
    \begin{equation*}
        y = \frac{X}{\Delta V}, \quad \Delta V = V_\mathrm{th}-E_\mathrm{rest}, \quad s = \frac{t}{\tau}, \quad \tau = \frac{1}{|Q'(0)|},
    \end{equation*}
producing a dimensionless drift series $\beta_i$ and a normalized input term. This yields a canonical form suitable for system-theoretic analysis and efficient simulation.

The same transformation procedure was applied to both the A-type potassium surrogate and the calcium surrogate polynomials.

\bibliography{pena}

\end{document}